\documentclass[pra,onecolumn,nofootinbib,superscriptaddress,notitlepage]{revtex4-2}

\usepackage{calrsfs}
\usepackage{ulem}
\usepackage{graphics}
\usepackage{epstopdf}
\usepackage{bm}
\usepackage{bbm}
\usepackage{amssymb}
\usepackage{epstopdf}
\usepackage{graphicx}
\usepackage[caption=false]{subfig}
\usepackage{amsmath}
\usepackage{color}
\usepackage{array}
\usepackage{lineno}

\begin{document}

\title{Spatial Mode Encoding for Quantum Key Distribution: From Hundreds to Thousands of Modes}

\author{Lukas \surname{Scarfe}}
\email{lscar039@uottawa.ca}
\thanks{These authors contributed equally to this work}
\affiliation{Nexus for Quantum Technologies, University of Ottawa, Ottawa ON Canada, K1N6N5}
\affiliation{National Research Council of Canada, 100 Sussex Drive, Ottawa ON Canada, K1A0R6}

\author{Yingwen \surname{Zhang}}
\email{yzhang6@uottawa.ca}
\thanks{These authors contributed equally to this work}
\affiliation{Nexus for Quantum Technologies, University of Ottawa, Ottawa ON Canada, K1N6N5}
\affiliation{National Research Council of Canada, 100 Sussex Drive, Ottawa ON Canada, K1A0R6}

\author{Ebrahim \surname{Karimi}}
\affiliation{Nexus for Quantum Technologies, University of Ottawa, Ottawa ON Canada, K1N6N5}
\affiliation{National Research Council of Canada, 100 Sussex Drive, Ottawa ON Canada, K1A0R6}
\affiliation{Institute for Quantum Studies, Chapman University, Orange, California 92866, USA}

\begin{abstract}
Here, we present a proof-of-principle high-dimensional quantum key distribution (QKD) protocol utilizing the position and momentum entanglement of photon pairs. The protocol exploits the fact that position and momentum form mutually unbiased bases, linked via a Fourier transform. One photon of the entangled pair is measured by the sender in a randomly chosen basis—either position or momentum—selected passively via a beam splitter. This projective measurement remotely prepares the partner photon in a corresponding spatial mode, which is sent to the receiver, who similarly performs a random measurement in one of the two bases. In this implementation, we achieve a photon information efficiency of 5.07 bits per photon using 90 spatial modes, and a maximum bit rate of 0.9\,Kb/s with 361 modes. To assess the scalability of this spatial-mode encoding scheme, we theoretically show that using a brighter entangled photon source along with next-generation single-photon cameras - featuring improved quantum efficiency, timing and spatial resolution - this approach could achieve 9 bits per photon at 2000 spatial modes, and a bit rate of over 700\,Mb/s at 4400 modes while accounting for finite-key effects. These results quantify the opportunities and performance bounds of spatially encoded, entanglement-based QKD and provide a benchmark for future high-dimensional quantum communication systems.
\end{abstract}
\maketitle

\section{Introduction}
Entanglement between two or more particles is one of the most profound concepts in quantum theory, forming the cornerstone of both foundational investigations and emerging quantum technologies~\cite{Entanglement}. On a conceptual level, entanglement challenges our classical understanding of nature, raising fundamental questions about the nature of reality and the principle of locality. From a technological perspective, entanglement serves as the primary resource, enabling capabilities beyond those of classical systems, including quantum computing~\cite{AghaeeRad2025,TensorionQC}, advanced sensing techniques~\cite{Pirandola2018,coldatomsensing,CVquantumsensing}, quantum teleportation~\cite{Sun2016,Bouwmeester1997}, and secure communication protocols~\cite{BENNETT20147,Li2025}.

Of the aforementioned technologies, quantum communications, in particular quantum key distribution (QKD), is primed to be one of the first quantum technologies to reach maturity and be implemented in wide-scale deployments ~\cite{Stanley_2022}. By generating a private key that is known to be secure, QKD allows individuals to communicate with absolute certainty of privacy, provided that the observed error rate in the quantum communication channel remains below a known threshold. While the field of quantum cryptography is broad, especially in recent years with different types of protocols being invented, two of the most mature QKD protocols schemes are prepare-and-measure based ~\cite{SISODIA2023100184} and entanglement ~\cite{EntanglespaceQKD}. In the former, one party Alice prepares random quantum states using a quantum random number generator and sends them to Bob. Entanglement-based protocols have both parties receive correlated photons from a shared entangled source. This process offers intrinsic randomness through the generation of entangled photons, most often through spontaneous parametric down-conversion (SPDC), a random process where the output photon pair can be entangled in multiple degrees of freedom~\cite{SPDC1}. With entanglement-based QKD it is possible to place the photon source at a trusted third party in between Alice and Bob allowing doubling the distance over which keys can be exchanged~\cite{DoubleDQKD,DIQKD}.

QKD, when implemented in high dimensions, can allow for higher information density per photon, as well as an increased error tolerance~\cite{HDQKD,HDQKDErr,dlevelQKD}, but this is often limited by experimental challenges in the efficient generation and detection of high-dimensional modes. Although high-dimensional QKD research has received growing attention~\cite{Sit17,bouchard2018experimental,Scarfe2025,zhong2015photon,Lib2025,Bhat2025}, practical systems still favor $2$-dimensional qubit-based  protocols using time or polarization modes due to their simplicity~\cite{Dynes:12,Liao2017}.

In this work, we present a proof-of-principle high-dimensional QKD scheme using position-momentum modes of entangled photons generated via SPDC. Here Alice detects one photon of an entangled pair, which is randomly directed via a 50:50 beam splitter into one of two mutually unbiased bases (MUBs): the position or momentum basis, which are conjugate variables linked by a Fourier transform~\cite{P-x_QKD}. This measurement projects the partner photon into a corresponding high-dimensional spatial mode, which is sent to Bob. Bob uses an identical detection setup, also randomly selecting between the two MUBs. In this way, high-dimensional state preparation and measurement are unified into a single passive process that eliminates the need for complex optical modulation or external random number generation, as both the MUB choice and photon-pair mode selection arise intrinsically from the SPDC process. Spatial mode detection is performed using event-based single-photon cameras~\cite{Nomerotski2019,ASI2}, which can time-tag every detected photon with nanosecond precision allowing for high-resolution coincidence imaging in both position and momentum bases, with the potential to access thousands of spatial modes. In the present experiment, we achieved a photon information efficiency of 5.07 bits per detected photon using 90 spatial modes and a maximum sifted key rate of 0.9\,Kb/s with 361 modes.  The performance of our current demonstration is significantly limited by the quantum efficiency, spatial and timing resolution of the camera as well as the brightness of the SPDC source. To assess the attainable performance under improved hardware conditions, we model the system using parameters consistent with next-generation superconducting nanowire array cameras~\cite{Wollman2019,Oripov2023} and increased source brightness. Under these assumptions, photon efficiencies approaching 9 bits per photon at 2000 spatial modes and secret-key rates exceeding 700\,Mb/s at 4400 modes are projected, providing an estimate of the achievable scaling of spatial-mode QKD with future detector advancements.

\section{Results}
\subsection{Experimental concept}
\begin{figure}
	\begin{center}
		\includegraphics[width=1\textwidth]{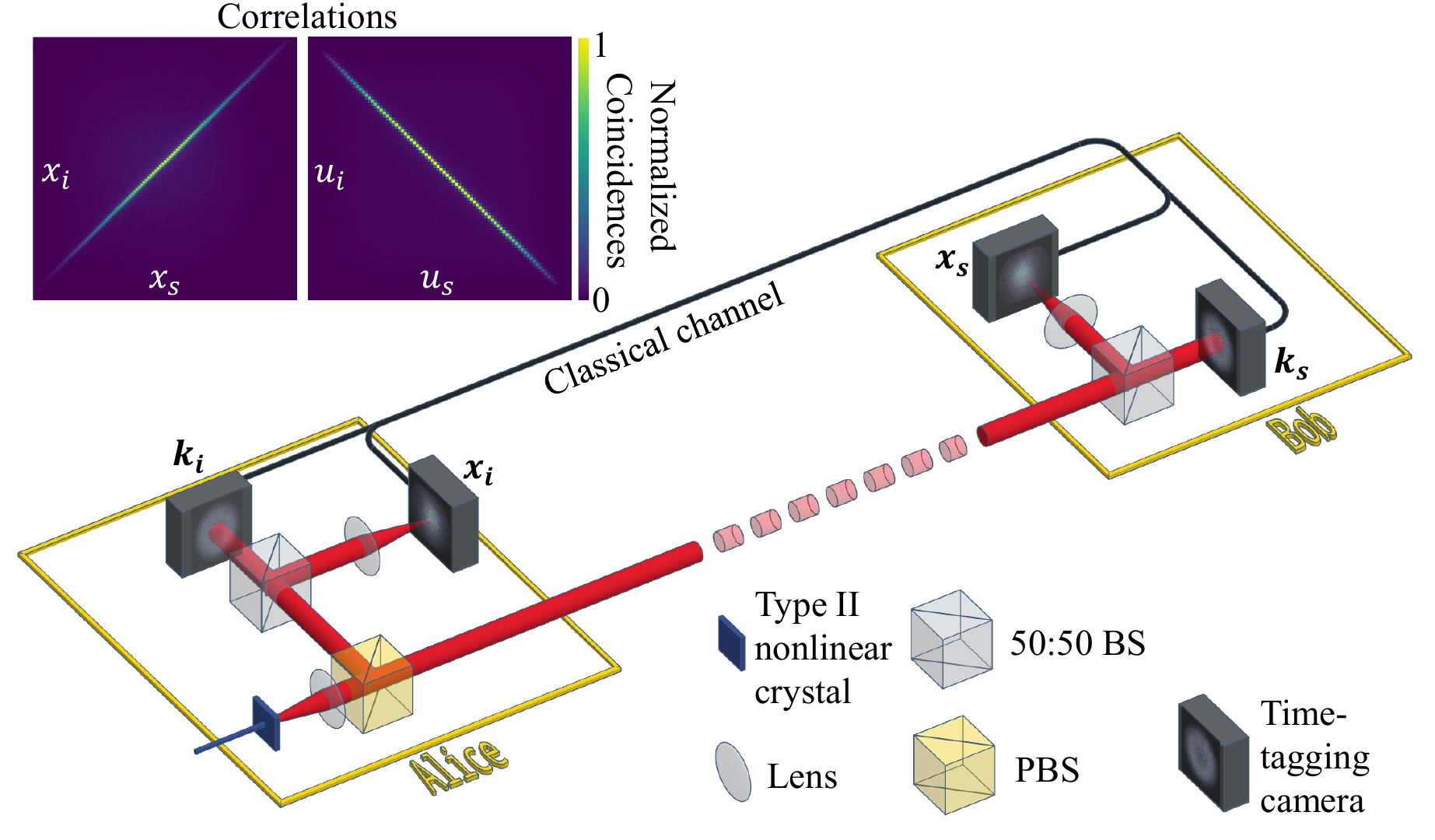}
		\caption{\textbf{Conceptual setup for position-momentum QKD}. Position-momentum entangled photon pairs with orthogonal polarization are created via Type-II SPDC by Alice who keeps the vertically polarized idler photon locally and sends the horizontally polarized signal photon to ``Bob". For detection at the two parties, the photons are randomly split to be measured in one of the two MUBs, either in position or momentum, by time-tagging cameras. Finally, the two parties compares their measurement bases via a classical channel to create their secret key. Inset on the top left shows the measured position and momentum correlations in the horizontal $x$ and $u$ direction, correlations in the vertical $y$ and $v$ directions looks near identical. Note that due to possessing only a single camera, in our experiment we subdivided the camera into four regions to act as the four cameras. See Supplementary Material for further information.}
		\label{fig:setup}
	\end{center}
\end{figure}

The conceptual setup for high-dimensional QKD using position–momentum entangled photons is shown in Fig.~\ref{fig:setup}. Photon pairs entangled in position and momentum are generated via Type-II spontaneous parametric down-conversion (SPDC) by Alice, producing orthogonally polarized photons. A polarizing beam splitter (PBS) separates the pair: the vertically polarized idler photon is retained by Alice, while the horizontally polarized signal photon is sent to Bob. At both ends, measurements are performed in one of two mutually unbiased bases (MUBs) — position $\bm{r}=(x,y)$ or momentum $\bm{k}=(u,v)$ — chosen at random using 50:50 beam splitters.  The act of measurement by Alice in a given basis projects Bob’s photon into a corresponding high-dimensional spatial mode in that same basis, effectively preparing the state that Bob receives. Time-tagging single-photon cameras enable spatially resolved coincidence detection of the photon pair in each basis. In our system, we measured a momentum correlation width of $\delta_k = 6.5 \times 10^{-3}$\,$\mu$m$^{-1}$ and a position correlation width of $\delta_r = 14$\,$\mu$m (Fig.~\ref{fig:setup} inset). The measured singles rates were approximately 300\,kHz in each of the four beams (signal and idler in both position and momentum configurations), with coincidence rates of about 5\,kHz between the signal and idler in both bases. All results reported in the following sections are based on 100 seconds of data acquisition.

After measurement, Alice and Bob publicly compare their basis choices and retain only the events measured in matching bases. The resulting sifted data can then be processed using standard error correction and privacy amplification procedures to extract a shared secret key. In the present work, we focus on the characterization of the high-dimensional encoding and the achievable photon information efficiency, leaving a full implementation of classical post-processing to future studies. 

This protocol benefits from fully passive state preparation and measurement. The spatial mode of each photon is intrinsically random due to the position–momentum entanglement of the SPDC source, requiring no external random number generation. Similarly, the measurement basis—either position or momentum—is selected passively via the 50:50 beam splitter, routing incoming photons to one of two mutually unbiased basis measurements. This fully passive implementation simplifies the experimental architecture while preserving security. Although the source is located with Alice in the conceptual setup, the protocol is compatible with a symmetric configuration where the source is placed at a trusted node between users, as in standard entanglement-based QKD. 

\subsection{High-dimensional performance and scaling}

The simplest entanglement-based QKD protocol to implement in high dimensions is the generalized BB84 protocol~\cite{BENNETT20147}, commonly referred to as BBM92 when using entangled photon sources~\cite{BBM92}. This protocol employs two mutually unbiased bases (MUBs), each consisting of $d$ orthogonal modes that define the dimensionality of the encoding space. When Alice and Bob measure their respective photons in the same basis, entanglement ensures their outcomes are ideally correlated. Any deviation from this correlation indicates errors, which may arise from noise or potential eavesdropping. By monitoring the error rate and ensuring it remains below a known threshold, Alice and Bob can confirm the security of their shared key. Outcomes measured in different bases are discarded during the sifting process, as they do not contribute to secure key generation. 

High-dimensional QKD protocols exhibit increased noise tolerance as the dimensionality $d$ grows. In the asymptotic limit of infinitely long keys, the net secret information per detected photon after sifting is given by~\cite{HDQKDSecure}:
\begin{equation}
    R = \log_2(d) - 2H_d(e),
    \label{eq:Rate}
\end{equation}
where $e$ denotes the quantum dit error rate (QDER), and $H_d(e) = -e\log_2\left(\frac{e}{d-1}\right) - (1-e)\log_2(1-e)$ is the $d$-dimensional Shannon entropy. Setting $R(e) = 0$ determines the critical error threshold above which secret key generation is no longer possible.

\begin{figure}
	\begin{center}
		\includegraphics[width=0.7\textwidth]{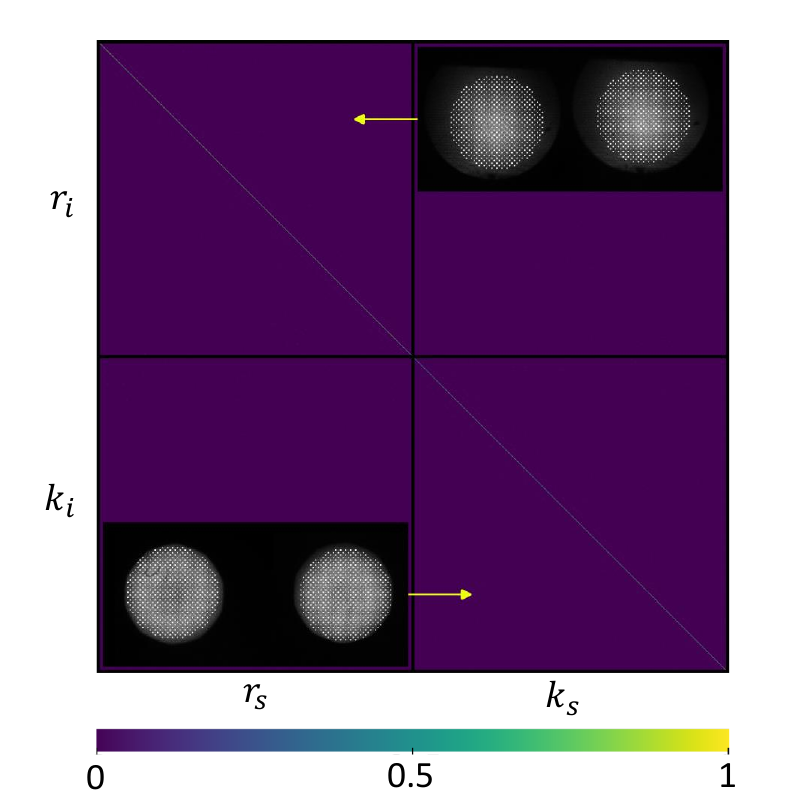}
		\caption{\textbf{Joint detection matrix $C_{r,k}$ for $d=545$.} $C_{r,k}$ exhibits a clear diagonal structure, indicating strong correlations between Alice’s and Bob’s measurement outcomes when they choose the same basis—these events contribute to the shared secret key. In the ideal case with no errors $C_{r,k}$ would be an identity matrix. Insets show the selected pixel layouts for the position $x$ and momentum $k$ beams, with Alice's measured beam on the left and Bob's on the right.}
		\label{fig:crosstalk}
	\end{center}
\end{figure}

To evaluate the QDER, the joint probability distribution of Alice’s and Bob’s outcomes when measuring in the same basis must be evaluated. This is represented by the joint detection matrix $C_{r,k}$, which gives the probability that Bob detects a particular mode given Alice's measurement result. An example of the measured $C_{r,k}$ for $d=545$ is shown in Fig.\,\ref{fig:crosstalk}. The QDER is then computed as:
\begin{equation}
    e = 1 - \frac{1}{d}\text{Tr}(C_{r,k}).
\end{equation}

$C_{r,k}$ can be determined theoretically through the position-momentum correlation function of SPDC and the coincidence background, 
this can be written as 
\begin{align}
    C_{r,k} &= A\left[\alpha_r(\bm{r}_s, \bm{r}_i) + \alpha_k(\bm{k}_s,\bm{k}_i) \right.\nonumber\\
    & \quad\left.+ \beta_r(\bm{r}_s, \bm{r}_i) + \beta_k(\bm{k}_s,\bm{k}_i)\right],
    \label{CrsTalkM}
\end{align}
where $A$ is a normalization constant. 

The $\alpha_\chi(\bm{\chi}_s, \bm{\chi}_i)$ terms contains the spatial correlation information and is given by
\begin{equation}
    \alpha_\chi(\bm{\chi}_s, \bm{\chi}_i) = \eta_s\eta_i P|\psi({\bm{\chi}}_s,{\bm{\chi}}_i)|^2,
\end{equation}
with $P$ the total SPDC pair rate in the position and momentum planes (assumed to be equal), $\eta_s$ and $\eta_i$ are the system detection efficiency for the signal and idler photons, and $\psi({\bm{\chi}}_s,{\bm{\chi}}_i)$ is the position or momentum correlation function.

The $\beta_\chi(\bm{\chi}_s, \bm{\chi}_i)$ terms are the background from accidental coincidence detection between uncorrelated photons. To second order, this is given by
\begin{equation}
    \beta_\chi(\bm{\chi}_s, \bm{\chi}_i) = \eta_s\eta_i \tau[p(\bm{\chi}_s)+b(\bm{\chi}_s)][p(\bm{\chi}_i)+b(\bm{\chi}_i)],
\end{equation}
where $\tau$ is the coincidence gating time, $p(\bm{\chi})$ is the SPDC pair rate at position $\bm{r}$ or momentum $\bm{k}$, and is related to the total pair rate as $p(\bm{\chi}_s) = P|\int\psi({\bm{\chi}}_s,{\bm{\chi}}_i)d^2\chi_i|^2$. Lastly, $b(\bm{\chi})$ is the background photon rate at the respective position or momentum. Refer to the Supplementary Materials for more details. 

In practical implementations, the spatial intensity distribution in both the position and momentum bases plays an important role, as non-uniform detection probabilities can introduce basis-dependent biases that must be accounted for in a complete security analysis. Such effects can be addressed at the classical post-processing stage, for example by appropriately weighting or selectively filtering detection events to balance the mode statistics, albeit at the expense of reduced raw key rates. Alternatively, tailoring the spatial profile of the pump beam to produce a more uniform SPDC intensity distribution in both conjugate planes offers a hardware-level solution that more efficiently utilizes the available spatial modes. In the present proof-of-principle study, we focus on characterizing the accessible dimensionality and photon information efficiency, while these optimization strategies provide clear avenues for further refinement.

\begin{figure}
	\begin{center}
		\includegraphics[width=0.6\textwidth]{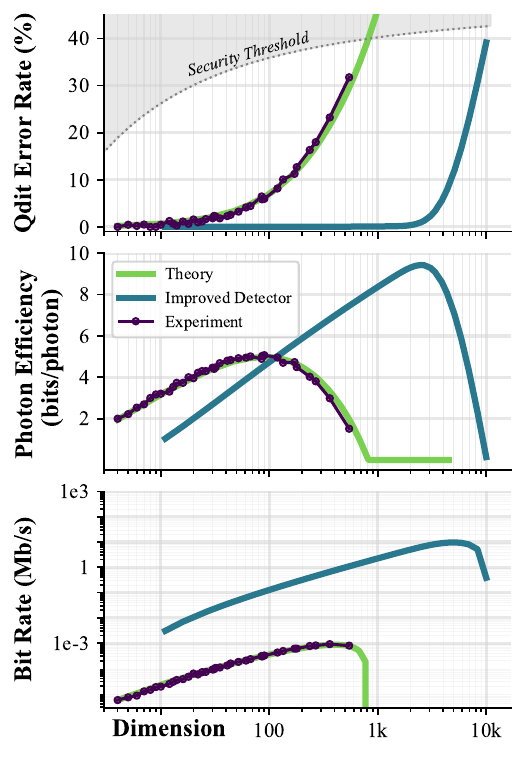}
		\caption{\textbf{Effect of increased dimensionality on the sifted key rate and the Qdit error rate (QDER).} Plot of the QDER (top), Photon efficiency (middle) and Bit rate (bottom) vs Dimension. The maximum error rate at which QKD could theoretically be implemented for each dimension is shown in the gray area of the top QDER plot and is denoted as the ``Security Threshold". The theoretical performance with current experimental parameters is shown as the green curve and compared to the experimentally measured performance, shown in purple. The cyan curve is the expected performance under finite key length while using next generation detectors as well as having a 90:10 beam splitter for MUB selection instead of 50:50.}
		\label{fig:KeyErr}
	\end{center}
\end{figure}

Figure~\ref{fig:KeyErr} presents the performance of high-dimensional QKD using position-momentum modes, for Hilbert space dimensions ranging from $d = 4$ to $10,000$, both experimentally and through theoretical modeling based on measured experimental parameters using Eq.\,\eqref{CrsTalkM}. The dimensionality is determined by the number of detection pixels and their spacing across the four correlated beams (details on mode selection and ordering are provided in the Supplementary Materials). The highest photon information efficiency is observed at $d = 90$, yielding 5.07 bits per detected photon, while the maximum raw bit rate is achieved at $d = 361$, reaching 0.9\,Kb/s.

In this proof-of-concept implementation, system performance is primarily constrained by the characteristics of the single-photon camera. With a camera quantum efficiency of approximately $8\%$~\cite{Vidyapin2022} the overall system efficiency is only $\sim2\%$. The available spatial resolution ($256\times256$) limits the ability to fully resolve the position–momentum correlations, and the timing resolution of approximately 8,ns~\cite{Vidyapin2022} restricts coincidence discrimination. Together, these factors reduce the achievable coincidence rate and increase background contributions. However, emerging technologies, particularly superconducting nanowire single-photon cameras~\cite{Wollman2019,Oripov2023}, are rapidly advancing, with expectations of $>90\%$ quantum efficiency, picosecond timing resolution and mega-pixel spatial resolution in the near future.

Figure~\ref{fig:KeyErr} also shows the projected theoretical QKD performance assuming next-generation detector parameters while retaining the same SPDC source used in the current experiment. In addition, we consider an asymmetric 90:10 beam splitter in place of the 50:50 configuration to increase encoding efficiency~\cite{Lo2005Efficient}. The model assume a total system efficiency of $\sim60\%$, a coincidence window of 100\,ps and a 1.5 times improved spatial resolution. In order to better model practical implementations, finite-key effects are also considered which requires modifying Eq.\,\eqref{eq:Rate} to~\cite{Yu2025}:
\begin{align}
    R=&\underset{{\beta \in (0,\frac{\varepsilon_{\text{sec}}}{2}]}}{\text{max}}\lfloor n_g \Big[ \log_2(d) -H_d\big(e_k+\delta(n_g,n_e,\beta)\big) \nonumber\\
    & - 1.2H_d(e_r)\Big]-\log_2\frac{8}{\beta^4\varepsilon_\text{cor}} \rfloor,
    \label{eq:finite_key_rate}
\end{align}
where $n_g$ and $n_e$ denote the numbers of bits used for raw key generation and parameter estimation, respectively. The quantities $e_r$, $e_k$ represent the error rates measured in the position and momentum bases. $\varepsilon_\text{sec}$, and $\varepsilon_\text{cor}$ represent chosen security parameters for the protocol ~\cite{Muller-Quade_2009}, and $\beta$ is an optimization parameter~\cite{Tomamichel2013} (See Supplementary Material for additional details of the finite-key analysis). 

Under these conditions, the photon information efficiency would increase to $\sim9$ bits per photon at $d \approx 2000$, and the corresponding bit rate could reach $\sim10$\,Mb/s with $d\approx 4400$.

\begin{figure}
	\begin{center}
		\includegraphics[width=0.6\textwidth]{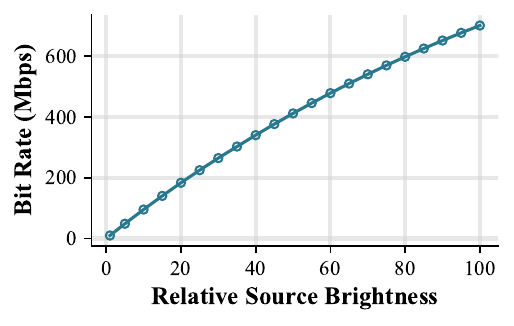}
		\caption{\textbf{Effect of increased source brightness on the sifted key rate} The expected sifted key rate in the finite-key regime with increased source brightness using next generation detectors.}
		\label{fig:brightness}
	\end{center}
\end{figure}

Further performance gains can be achieved by increasing the brightness of the SPDC source. The present experiment employs a 40\,mW pump and a 1\,mm-thick Type-II ppKTP crystal. Replacing this configuration with a higher-brightness implementation — such as Type-0 phase matching, increased pump power, and a thicker nonlinear crystal — would near proportionally enhance the coincidence rate and, consequently, the achievable key rate. As illustrated in Fig.~\ref{fig:brightness}, a 100-fold increase in SPDC brightness with next generation detectors would yield a projected key rates of over 700\,Mb/s at $d\approx4400$. To achieve this, transitioning from Type-II to Type-0 SPDC alone can provide more than a 50-fold increase in photon pair generation efficiency~\cite{Steinlechner2014}. Implementing this modification requires only a simple adjustment to the experimental setup, namely replacing the polarizing beam splitter currently used to separate photon pairs between Alice and Bob with a dichroic mirror centered on the degenerate wavelength.

\section{Discussion}

In conclusion, we have demonstrated a proof-of-concept experimental realization of high-dimensional QKD using position–momentum entangled photons. We achieved a maximum photon information efficiency of 5.07 bits per photon with 90 spatial modes and a maximum raw bit rate of 0.9\,kb/s with 361 modes. Our theoretical model, in good agreement with the experimental results, predicts that with next-generation superconducting nanowire array cameras and a brighter SPDC source could increase photon efficiencies to approximately 9 bits per photon at 2000 spatial modes and achieve bit rates exceeding 700\,Mb/s at 4400 spatial modes. Our results indicate that the implementation of future detectors will potentially allow for spatial-mode encoding to be used as a realistic basis for QKD in real-world systems. We expect this performance can potentially be further increased by multiplexing with additional degree of freedom such as polarization and time-bins~\cite{zhong2015photon,TimebinHDQKD}.

Beyond free-space implementations, spatial-mode QKD could be potentially integrated into fiber-based quantum networks using multimode fibers. By combining spatial encoding with time-bin or polarization encoding, the information capacity of the quantum channel can be further increased. Realizing this requires compensation techniques to correct for beam distortions in the fiber~\cite{Zhou2021,Abdulaziz2023}, but the combination of multiple degrees of freedom offers a promising route toward ultra-high-dimensional QKD. 

In principle, the approach demonstrated here can be extended to other spatial mode bases, such as Laguerre–Gauss (LG) and Hermite–Gauss (HG) modes, which offer shape-invariant properties and well-established mode-sorting techniques~\cite{PhysRevLett.105.153601,Lavery:12,Larocque:17,fontaine2019laguerre}. These modes provide a robust and scalable framework for high-dimensional quantum communication~\cite{Mirhosseini_2015}, but efficient generation and detection in high dimensions are typically more challenging than in the pixel basis. Our technique, which leverages entanglement-based projection to prepare high-dimensional modes, could serve as a foundation for rapidly generating and detecting LG and HG modes, potentially overcoming these challenges. A comparative study in the future of pixel-based versus analytically defined modes, evaluating channel capacity, error rates, and experimental feasibility, would be valuable for optimizing high-dimensional QKD implementations. 

\section{Methods}
\begin{figure*}
	\begin{center}
		\includegraphics[width=1\textwidth]{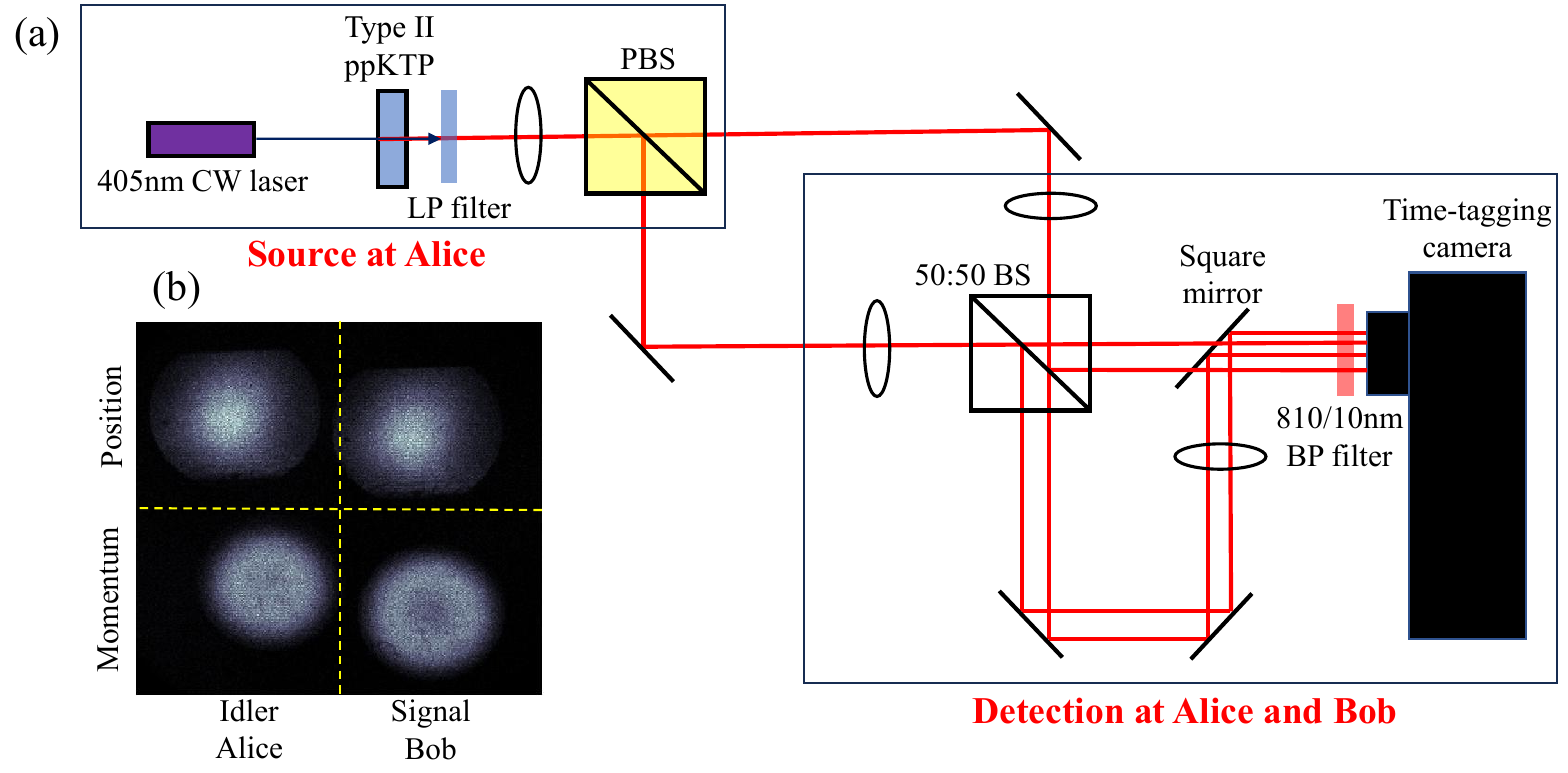}
		\caption{(a) Experimental setup for position-momentum QKD. LP-filter: long-pass filter, BP-filter: band-pass filter, PBS: polarizing beam-splitter, BS: beam-splitter (b) Image captured on camera of the position and momentum planes of SPDC.}
		\label{fig:setup}
	\end{center}
\end{figure*}
The experimental setup for demonstrating high-dimensional QKD using position-momentum entangled photons is shown in Figure~\ref{fig:setup}\,(a). Due to possessing only a single time-tagging camera (TPX3CAM)~\cite{Nomerotski2019,ASI2}, it is used as the detector for both Alice and Bob for this experimental demonstration. In practice, at least 2, preferably 4, such cameras should be used, one for each party (for 2 cameras) or one for each MUB at each party (for 4 cameras).

\noindent\textbf{Source at Alice}: Position-momentum entangled photon pairs with orthogonal polarization are generated at a rate of approximately $8\times 10^6$ photon pairs per second by pumping a 1\,mm thick Type-II ppKTP crystal with a 405\,nm CW laser having 40\,mW power and a collimated beam with a beam width of $0.48$\,mm at the crystal plane. The orthogonally polarized photons are separated using a polarizing beam-splitter (PBS), with one photon kept by Alice and the partner sent to Bob. 

\noindent\textbf{Detection at Alice and Bob}: The photons enter a 50:50 beamsplitter which will allow each photon to have a 50\% probability of being detected in one of the two MUBs, position or momentum. A square mirror is used to recombine the four beams onto the camera, where the two beams to be measured in the position plane pass over the mirror unaffected, and the two beams to be measured in the momentum plane are reflected by the mirror onto the camera. As seen in Figure ~\ref{fig:setup}\,(b), the camera sensors are split into four quadrants, with the left two quadrants used as the detector for Alice's MUBS and the right two quadrants used as the detector for Bob's MUBS. 
Imaging lenses were used to image the near-field (position plane) of the ppKTP crystal onto the camera with a magnification of $\sim 5$ times and image the far-field (momentum plane) onto the camera with a demagnification of $\sim 5$ times.
A singles rate of $\sim 300$\,kHz in each of the four beams and a coincidence rate of $\sim5$\,kHz were measured between the signal and idler NF and also the signal and idler FF. Together, this gives a total detection efficiency of $\sim 2$\% for the experimental setup (see more details on this in section III), which accounts for the $8$\% detection efficiency of the TPX3CAM~\cite{Vidyapin2022}, $50\%$ loss from the 50:50 split by the PBS and losses from the optics in general, details on data processing of the raw TPX3CAM data can be found in \cite{Zhao2017,Vidyapin2022}.

\section*{Acknowledgements}
The authors wish to express their sincere gratitude to Dr. Alessio D’Errico for his insightful discussions, valuable feedback and comments on this work. This work was supported by Canada Research Chairs; National Research Council of Canada High-Throughput and Secure Networks (HTSN) Challenge Program; and the Alliance Consortia Quantum Grant (QUINT, ARAQNE).

\section*{Appendix}
\subsection{Position-momentum correlation function}
In the low gain regime, the position-momentum entangled state of SPDC in transverse momentum space can be written as
\begin{equation}
    |\Psi\rangle = \int\int \phi(\bm{k}_s,\bm{k}_i) |\bm{k}_s,\bm{k}_i\rangle\, d^2k_s d^2k_i,
\end{equation}
and the biphoton wavefunction $\phi(\bm{k}_s,\bm{k}_i)$ can be approximated as a double-Gaussian approximation~\cite{Law2004,Chan2007,Schneeloch2016}
\begin{align}
    \phi({\bm{k}}_s,{\bm{k}}_i) &\propto \exp\left(\frac{-|\bm{k}_s-\bm{k}_i|^2}{2\sigma_k^2}\right)\nonumber\\
    &\quad \times\exp\left(\frac{-|\bm{k}_s+\bm{k}_i|^2}{2\delta_k^2}\right),
    \label{kcorr}
\end{align}
and in position space, the biphoton wavefunction is
\begin{align}
    \psi({\bm{r}}_s,{\bm{r}}_i) &\propto  \exp\left(\frac{-|\bm{r}_s-\bm{r}_i|^2}{2\delta_r^2}\right)\nonumber\\
    &\quad\times\exp\left(\frac{-|\bm{r}_s+\bm{r}_i|^2}{2\sigma_r^2}\right).
    \label{rcorr}
\end{align}

In theory, $\delta_k \approx 1/(2\sigma_p)$; $\delta_r\approx\sqrt{\frac{2\alpha L \lambda_p}{\pi}}$, and related by a Fourier transform, $\sigma_k = 2/\delta_r$; $\sigma_r = 1/2\delta_k$, with $\sigma_p$ being the pump beam width, $L$ the crystal length, $\lambda_p$ the pump wavelength, and $\alpha = 0.455$ is a constant factor from the Gaussian approximation of the sinc phase matching function~\cite{Chan2007}. Using the experimental parameters $\sigma_p = 0.48$\,mm, $L = 1$\,mm and $\lambda_p = 405$\,nm, this gives an expected value of $\delta_r = 11\,\mu$m, $\sigma_r = 500\,\mu$m, $\delta_k = 1.0\times10^{-3}\,\mu$m$^{-1}$ and $\sigma_k = 0.18\,\mu$m$^{-1}$ 

\begin{figure}
	\begin{center}
		\includegraphics[width=0.8\textwidth]{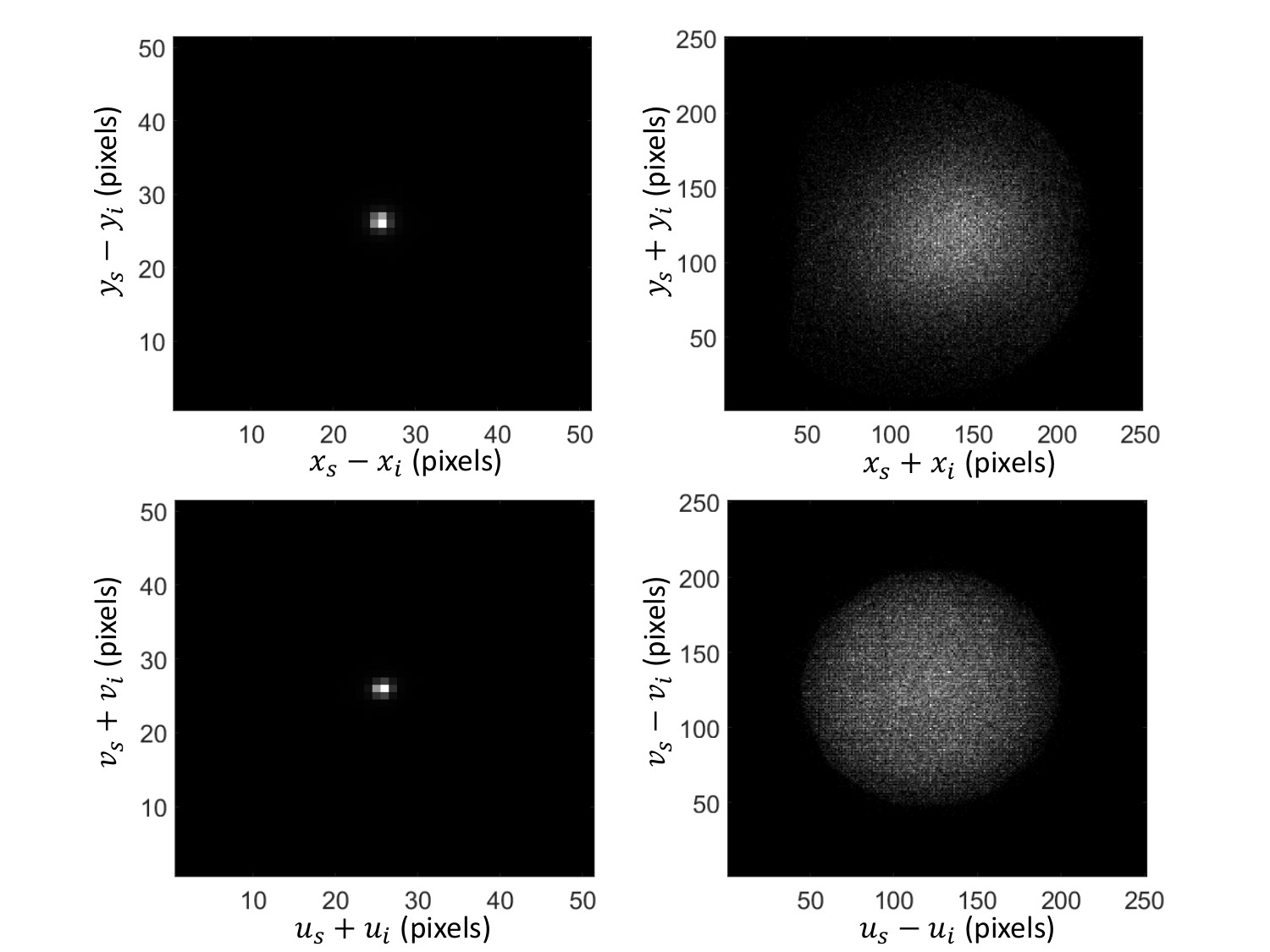}
		\caption{Plot of position and momentum intensity correlations. Fitted gaussian width of $\bm{r}_s-\bm{r}_i$ correlation is 0.90 pixels, $\bm{r}_s+\bm{r}_i$ correlation is 36.8 pixels, $\bm{k}_s+\bm{k}_i$ correlation is 0.65 pixels and $\bm{k}_s-\bm{k}_i$ correlation is 34.5 pixels.}
		\label{Supp2}
	\end{center}
\end{figure}

Figure~\ref{Supp2} shows the measured $\bm{r}_s-\bm{r}_i$, $\bm{r}_s+\bm{r}_i$, $\bm{k}_s-\bm{k}_i$, $\bm{k}_s+\bm{k}_i$ correlations. Fitting a gaussian gives a width of 0.90 pixels for the $\bm{r}_s-\bm{r}_i$ correlation, 36.8 pixels for $\bm{r}_s+\bm{r}_i$ correlation, 0.65 pixels for $\bm{k}_s+\bm{k}_i$ correlation and 34.5 pixels for $\bm{k}_s-\bm{k}_i$ correlation. Converting the beam width from an intensity measurement to amplitude will require multiplication by a factor of $\sqrt{2}$ and using the pixel pitch of 55\,$\mu$m and a magnification factor of 5 gives $\delta_r = 14\,\mu$m,  $\sigma_r = 573\,\mu$m. Converting from position space to momentum space using the wave number $k=2\pi/\lambda$ and a lens of focal length $300$\,mm gives $\delta_k = 6.5\times10^{-3}\,\mu$m$^{-1}$, $\sigma_k = 0.34\,\mu$m$^{-1}$.

The large discrepancy between the expected and measured $\delta_k$  is mainly due to the limited camera resolution being unable to fully resolve the correlation. Thus, for the theoretical model discussed in the next section, we will be using the measured correlation values instead of the expected values.

\subsection{Finite-key security bound for position-momentum modes entangled qudit QKD}

Here we provide the mathematical framework to a bound secret key length and its affect on the key rate $R$. We base our analysis on the work done on a similar protocol for ququarts in a time-bin based BBM92 protocol~\cite{Yu2025}. A QKD protocol can be considered $\varepsilon$-secure if it is both $\varepsilon_{\text{cor}}$ (correct) and $\varepsilon_{\text{sec}}$ (secret). A protocol is $\varepsilon_{\text{cor}}$-correct if the final-key bit strings of Alice ($S_A$) and Bob ($S_B$), are identical in nearly all cases, where the probability that they are different is $\varepsilon_{\text{cor}}$. The protocol is $\varepsilon_{\text{sec}}$-secret if the final key held by either Alice or Bob is uniformly distributed and independent of the the system that is held by a hypothetical eavesdropper. In essence, $\varepsilon_{\text{sec}}$ is the probability that Eve is able to extract useful information~\cite{Muller-Quade_2009}.

Using the quantum leftover-hash lemma~\cite{Tomamichel2013}, the boundary of the secret key length that can be extracted from Alice's raw key ($X$) given Eves total information on the system($E$) is given by:

\begin{align}
    l=\underset{{\beta \in (0,\frac{\varepsilon_{\text{sec}}}{2}]}}{\text{max}}=\lfloor H_{\text{min}}^{\frac{\varepsilon_{\text{sec}}}{2}-\beta} (X|E) +4 \log_2\beta -2 \rfloor
    \label{secret_key_length_raw}
\end{align}

Here, $H_{\text{min}}^{\frac{\varepsilon_{\text{sec}}}{2}-\beta}$ is the smooth minimum-entropy of $X$ given $E$, using an optimization parameter ($\beta$)~\cite{Tomamichel2013}. The bound for the smooth minimum-entropy is given by
\begin{align}
    H_{\text{min}}^{\frac{\varepsilon_{\text{sec}}}{2}-\beta} (X|E) \geq & n_g \Big[ \log_2(d) -H_d\big(e_k+\delta(n_g,n_e,\beta)\big)\Big]\nonumber\\
    & - \text{Leak}_\text{EC}(e_r)-\log_2\frac{2}{\varepsilon_\text{cor}},
    \label{smooth_min_entropy}
\end{align}
where the amount of information leaked during error correction is given by $\text{Leak}_\text{EC}(e_r)=1.2nH_d(e_r)$ and $e_r,e_k$ are the measured position and momentum error rates respectively. Additionally $\log_2\frac{2}{\varepsilon_\text{cor}}$ is the number of bits published for error verification, and $H_d(e) = -e\log_2\left(\frac{e}{d-1}\right) - (1-e)\log_2(1-e)$ is the $d$-dimensional Shannon entropy.

Due to the finite-key statistics, we must account for the fact that the subset of bits used for the error estimation does not necessarily perfectly represent the entire bit-string. This is done through the term $\delta(n_g,n_e,\beta)=\sqrt{(n_g+n_e)(n_e+1)\ln(2/\beta)/n_gn_e^2}$. This term disappears as both $n_g$ and $n_e$ approach infinity. $n_g$ and $n_e$ are the number of bits used for the raw key generation and the number of bits used for parameter estimation, respectively. Finally the secret key length is given by:

\begin{align}
    l=&\underset{{\beta \in (0,\frac{\varepsilon_{\text{sec}}}{2}]}}{\text{max}}\lfloor n_g \Big[ \log_2(d) -H_d\big(e_k+\delta(n_g,n_e,\beta)\big) \nonumber\\
    & - 1.2H_d(e_r)\Big]-\log_2\frac{8}{\beta^4\varepsilon_\text{cor}} \rfloor.
    \label{secret_key_length_final}
\end{align}

In our theoretical model we assume for simplicity of analysis that we generate a key once every second, allowing us to use a simple one-to-one conversion from key length to key rate $l=R$. We set $\varepsilon_\text{cor}=10^{-9}$ and $\varepsilon_\text{sec}=10^{-12}$.  We sacrifice $10\%$ of our measured photons to be used for parameter estimation, $n_e=\frac{1}{9}n_g$.

\subsection{Theoretical model of QKD with position-momentum modes}

The joint detection matrix $C_{r,k}$ for position-momentum mode QKD can be written as the sum of four parts, 
\begin{align}
    C_{r,k} &= A\left[\alpha_r(\bm{r}_s, \bm{r}_i) + \alpha_k(\bm{k}_s,\bm{k}_i) \right.\nonumber\\
    & \quad\left.+ \beta_r(\bm{r}_s, \bm{r}_i) + \beta_k(\bm{k}_s,\bm{k}_i)\right],
    \label{CrsTalkM}
\end{align}
where the first two $\alpha$ terms describes the position and momentum correlation property of the SPDC photons and the $\beta$ terms gives the accidental coincidences coming from uncorrelated SPDC and background photons. $A$ is a normalization constant.

Assuming a total SPDC pair rate of $P$, and a detection efficiency of $\eta_s$ and $\eta_i$ for the two photons, $\alpha(\bm{r}_s, \bm{r}_i,\bm{k}_s,\bm{k}_i)$ can be written as
\begin{align}
    \alpha_r(\bm{r}_s, \bm{r}_i) &= \eta_s\eta_i P|\psi({\bm{r}}_s,{\bm{r}}_i)|^2 \nonumber\\
    \alpha_k(\bm{k}_s, \bm{k}_i) &= \eta_s\eta_i P|\phi({\bm{k}}_s,{-\bm{k}}_i)|^2,
    \label{CrsTalkC}
\end{align}
where $\psi({\bm{r}}_s,{\bm{r}}_i)$ and $\phi({\bm{k}}_s,{-\bm{k}}_i)$ are the position and momentum correlation functions given by Eq.\eqref{rcorr} and Eq.\eqref{kcorr} respectively. Note that the sign of $\bm{k}_i$ has been reversed so the momentum anti-correlation relation is displayed as a correlation relation in $C_{r,k}$ and we have also assumed for simplicity that $\eta_s, \eta_i$ are the same for the two planes.

The accidental coincidence terms $\beta$ is simply the probability for two uncorrelated photons to be detected in coincidence between two locations, to second order this is given by
\begin{align}
    \beta_r(\bm{r}_s, \bm{r}_i) &= \eta_s\eta_i \tau[p(\bm{r}_s)+b(\bm{r}_s)][p(\bm{r}_i)+b(\bm{r}_i)]\nonumber\\
    \beta_k(\bm{k}_s, \bm{k}_i) &= \eta_s\eta_i \tau[p(\bm{k}_s)+b(\bm{k}_s)][p(\bm{k}_i)+b(\bm{k}_i)]
    \label{CrsTalkB}
\end{align}
where $\tau$ is the coincidence gating time, $p(\bm{r})$; $p(\bm{k})$ are the SPDC pair rate with position $\bm{r}$ or momentum $\bm{k}$, this is related to the total pair rate as $p(\bm{r}_s) = P|\int\psi({\bm{r}}_s,{\bm{r}}_i)d^2r_i|^2$, $p(\bm{r}_i) = P|\int\psi({\bm{r}}_s,{\bm{r}}_i)d^2r_s|^2$ and similarly for $p(\bm{k}_s)$ and $p(\bm{k}_i)$. $b(\bm{r})$, $b(\bm{k})$ are the background photon rate at the respective position or momentum which we will assume to be a constant $b(\bm{k}) = b(\bm{r}) = B/N$ with $B$ being the total number of background photons and $N$ the total number of pixels in the beam. 


Experimentally we have the measurement of the singles given by $\eta(P+B)\approx 3\times10^5$, assuming $\eta = \eta_s = \eta_i$. The total number of temporally correlated events between two beams is $\eta^2\left[P+\tau(P+B)^2\right] \approx 5000$, and lastly the total number of spatio-temporal correlated events between two beams is $\eta^2\left[P+\frac{\tau\delta}{N}(P+B)^2\right]\approx \eta^2P\approx 3300$, given that the spatial correlation width $\delta \ll N$. From this we can work out $P \approx 8.1 \times 10^6$, $B \approx 6.7 \times 10^6$ and $\eta \approx 0.02$ given that we used $\tau = 20$\,ns for this experiment.

The expected QKD performance based on Eq.\eqref{CrsTalkM}, \eqref{CrsTalkC}, \eqref{CrsTalkB} using the above parameters is shown in fig.\,3 of the main text, where a maximum photon efficiency of 5\,bits/photon at 90 modes and a maximum bit rate of 0.9\,Kb/s at 400 modes is achieved. 

\subsection{Mode Orderings}
To evaluate the impact of mode structure and spatial arrangement on the performance of a quantum key distribution (QKD) protocol, we performed a sub-sampling of the total number of pixels available in our detector. This approach allowed us to investigate how different spatial mode orderings and inter-mode spacing influence the system’s error rates and overall performance. Specifically, we examined three distinct configurations: (1) a Cartesian grid pattern aligned with the native pixel layout of the detector, (2) an angled-grid pattern rotated by $45^{\circ}$ relative to the detector’s orientation, and (3) a hexagonal layout, known for its efficient packing within circular apertures.

\begin{figure}[h]
	\begin{center}
		\includegraphics[width=1\textwidth]{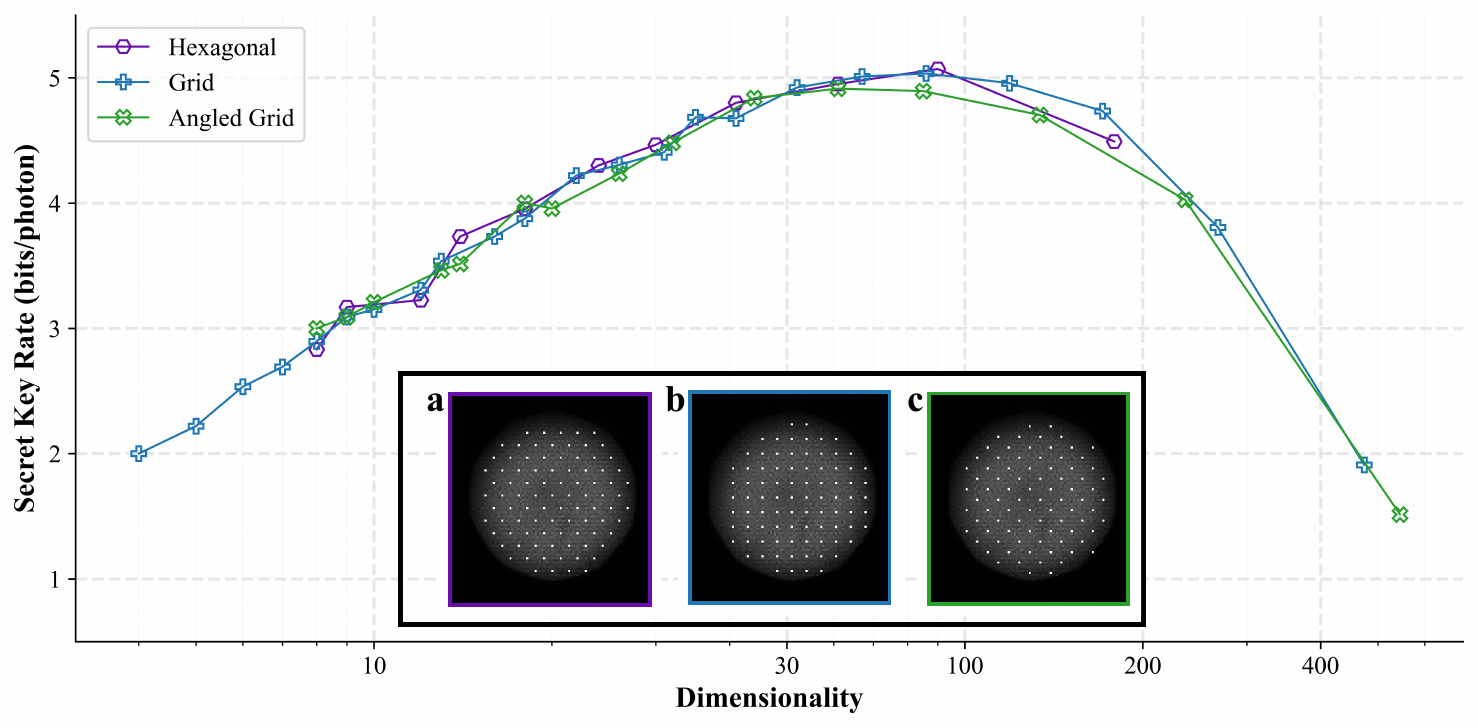}
		\caption{Secret key rate using different pixel mode orderings. Each illuminated pixel is chosen as a mode. Insets a, b, and c demonstrate the orderings of the hexagonal grid, the cartesian grid, and the angled cartesian grid, respectively. }
		\label{Supp4}
	\end{center}
\end{figure}

Contrary to our expectations, however, no particular layout demonstrated a clear advantage over the others. As shown in Fig.~\ref{Supp4}, all three configurations yielded comparable key rates when tested at similar dimensionalities. This suggests that, within the regime tested, the specific ordering of spatial modes does not significantly impact QKD performance, and other factors—such as inter-mode spacing or detector noise—may play a more critical role. By testing them all, we gained access to more dimensions than we would have otherwise by using a single configuration. These results were combined into a single dataset, where any overlapping dimensions used the mode ordering with the lowest error rate.

\subsection{Limitations of Individual Pixel Modes}
In our implementation, each spatial mode corresponds to an individual pixel on the detector. However, this approach utilizes only a small fraction of the detector’s active area. In the best case, just $12.7\%$ of the available $4293$ pixels are used as modes. As a result, when Alice and Bob choose opposite measurement bases, the probability of both photons landing on predefined mode pixels is low--with the probability proportional to $(d/4293)^2$ for any dimension $d$, assuming uncorrelated detection positions. Conversely, when both parties measure in the same basis, a photon detected on a mode pixel by Alice is highly likely to have its pair detected on the corresponding mode pixel by Bob.

As the protocol’s dimensionality increases, the spatial separation between neighboring modes decreases, leading to greater overlap and an increase in cross-talk errors. Consequently, the quantum dit error rate rises with dimension. Despite this, the photon information efficiency initially increases with dimensionality and peaks at $d = 90$, suggesting this to be the system’s optimal operating point in terms of balancing photon information density and error performance.

\bibliography{sample}

\end{document}